# Nonlinear vibrational modes in graphene: group-theoretical results

G.M. Chechin[1,2,†], D.S. Ryabov[2], S.A. Shcherbinin[2]

[†]gchechin@gmail.com

[1]Southern Federal University, Department of physics,

Zorge Str., 5, 344090, Rostov-on-Don, Russia

[2]Southern Federal University, Institute of physics,

Stachki Ave., 194, 344090, Rostov-on-Don, Russia

In-plane nonlinear delocalized vibrations in uniformly stretched single-layer graphene (space group *P6mm*) are considered with the aid of the group-theoretical methods. These methods were developed by authors earlier in the framework of the theory of the bushes of nonlinear normal modes (NNMs). We have found that only 4 symmetry-determined NNMs (one-dimensional bushes), as well as 14 two-dimensional, 1 three-dimensional and 6 four-dimensional vibrational bushes are possible in graphene. They are exact solutions to the dynamical equations of this two-dimensional crystal. Prospects of further research are discussed.

**Keywords**: lattice dynamics, nonlinear normal modes, anharmonic vibrations, group- theoretical methods, graphene.

## 1. Introduction

Various types of nonlinear vibrations in systems with discrete symmetry can exist. Discrete breathers (DBs) represent one type of such vibrations. They are spatially localized and time-periodic oscillations of the crystal lattice [1-4]. Another type of nonlinear lattice vibrations is bushes of nonlinear normal modes (NNMs). The concept of these dynamical objects was introduced in [5-7]. As well as discrete breathers, they are *exact* solutions to equations of motion. Each bush represents a set of delocalized NNMs which is conserved in time.

The number *m* of modes entering into a given bush defines its dimension. One-dimensional bushes represent individual nonlinear normal modes by Rosenberg [8]. Each Rosenberg mode describes a *periodic* dynamical regime. In [9-14], the bushes of vibrational modes were studied in various nonlinear systems of different physical nature with some point and space symmetry groups.

In microscopic systems, experimental investigation of bushes of vibrational modes and discrete breathers is associated with great difficulties. In view of this, ab initio computer simulations, based on the density functional theory (DFT) [15], are very important for studying properties of these dynamical objects. The efficiency of DFT methods for investigation of DBs in graphane and graphene was demonstrated in papers [16-18]. In [19], we apply DFT simulations for verification group-theoretical results of the bush theory using as an example the $SF_6$ molecule.

The present paper is devoted to studying low-dimensional bushes of nonlinear vibrational modes in uniformly stretched single-layer graphene (space group *P6mm*).

## 2. Nonlinear normal modes

We consider the delocalized in-plane vibrational modes in graphene. Conventional (linear) normal modes (LNMs) represent the simplest example of such dynamical regimes. Being introduced in the harmonic approximation, they are exact solutions to linear dynamical equations corresponding to this approximation. According to the well known Wigner theorem, LNMs can be classified by irreducible representations (irreps) of the group $G_0$ describing the symmetry of the considered Hamiltonian system in equilibrium state. In particular, this means that degrees of degeneracy of vibrational frequencies of molecules and crystals are determined by dimensions of these irreps, while their basis vectors provide us with the instant distributions of the displacements of all atoms in the considered system (atomic displacement patterns). Any displacement pattern possesses a certain symmetry group *G*, which is a *subgroup* of the group $G_0$ ($G \subseteq G_0$).

In the harmonic approximation, LNMs are independent from each other, while these modes begin to interact if one takes into account *weak* anharmonic terms (let us remember different types of phonon-phonon interactions in the framework of solid state physics). Obviously, the theory considering interactions between LNMs turns out to be only approximate. Therefore, the following question can be posed: "Are there some *exact* solutions *beyond* the harmonic approximation?"

Many years ago, Lyapunov proved the existence of periodic solutions in nonlinear systems which can be obtained by the continuation of the LNMs by the parameter describing the strengths of nonlinearity [20]. However, Lyapunov's continuation procedure usually can be fulfilled only in a very small vicinity of the LNM and practically it occurs rather cumbersome. As a consequence, nonlinear normal modes by Lyapunov are of minor importance in physics.

Another type of nonlinear normal modes was introduced by Rosenberg [8] (see also [21]). By definition, each Rosenberg NNM represents such dynamical regime in which all vibrational degrees of freedom $x_i(t)$ are proportional to the same function $f(t)$:

$$x_i(t) = a_i f(t), \qquad (1)$$

where $a_i$ are constant coefficients. Unlike Lyapunov's modes, NNMs by Rosenberg may exist in Hamiltonian systems with a high-degree nonlinearity. However, their existence is possible only in very specific dynamical systems. In particular, they can exist in mechanical systems, whose potential energy is a homogeneous function of all its arguments. However, it was shown in [9-11] that there are some symmetry-related reasons which cause the possibility of the Rosenberg modes existence for general type of the system potential energy. These modes we call symmetry-determined Rosenberg nonlinear normal modes (SD NNMs). Below, we discuss only such type of Rosenberg modes and, therefore, omit specification "symmetry-determined". It is essential that in a given dynamical system only a *finite number* of SD NNMs can exist. For example, it was proved in [10] that only six SD NNMs are possible in the FPU-β chains.

Each Rosenberg NNM corresponds to a certain *periodical* dynamical regime, depending on one parameter which is the amplitude of a given mode. In the limit of small amplitudes, each Rosenberg mode transforms to some LNM and the function $f(t)$ turns into the function $\cos(\omega t + \phi_0)$, where $\omega$ and $\phi_0$ are the frequency and initial phase of oscillation, respectively.

The general theory of nonlinear dynamical regimes in physical systems with discrete symmetries was developed in [5-7]. It is a theory of the *bushes* of nonlinear normal modes. In the framework of this theory, NNMs by Rosenberg represent one-dimensional bushes.

### 3. Bushes of nonlinear normal modes

Hereafter, speaking about excitation of a given dynamical regime, we mean that at the initial instant ($t = 0$) we set some atomic displacement pattern, and then allow the system to evolve freely in time. If exited nonlinear normal mode represents an exact solution to the dynamical equations of the physical system, it will be live infinitely long in time. However, we can encounter with essentially different situation. If the initially excited nonlinear normal mode is not an exact solution, it involves some other NNMs into the vibrational process. The *complete collection* of these modes corresponds to a certain exact solution and we call it *bush of NNMs* [5-7] (see also the review paper [22]). The number $m$ of modes entering the given bush conserves during time-evolution, while the mode amplitudes change in time. The number $m$ is the dimension of the bush.

Dynamics of *m*-dimensional bush is described by $m$ second-order differential equations of Newton type. Its Fourier spectrum contains $m$ basic frequencies, as well as their different integer linear combinations. Therefore, unlike individual nonlinear normal modes (bushes with $m = 1$), *m*-dimensional bush for $m > 1$ represents a *quasiperiodical* dynamical regime.

The total energy of the initial excitation turns out to be trapped in the given bush. NNM whose initial excitation generates the bush, we call *root* (primary) mode of this bush, while all its other NNMs – *secondary* modes. The symmetry group of the root mode determines that of the whole bush. Secondary modes possess the same or *higher* symmetry groups[*].

The mathematical apparatus of the bush theory is based on the irreducible representations of the parent symmetry group $G_0$ (this is the symmetry group of the system in equilibrium or that of its Hamiltonian). Below, we outline briefly the main ideas of the bush theory mathematics.

Firstly, let us note that initially the group-theoretical methods, underlying the theory of the bushes of NNMs, were applied to the delocalized dynamical regimes [5-7], while later the similar methods have been used for analyzing localized vibrations (discrete breathers and quasibreathers [25]) [26], as well as for studying chaotic oscillations [27].

All dynamical regimes in a given physical system with the parent symmetry group $G_0$ can be classified by its subgroups $G_j$ ($G_j \subseteq G_0$). As was already mentioned, the symmetry group $G_j$ of any dynamical regime is determined by the symmetry of its atomic displacement pattern $\boldsymbol{\delta}_j$. For any *stable* dynamical regime the group $G_j$ is conserved during the time evolution. Acting on $\boldsymbol{\delta}_j$ by all symmetry elements of the group $G_0$, we can select only those for which it is invariant. The full set of these elements forms the group $G_j$ ($G_j \subseteq G_0$). In particular, the group $G_j$ may be trivial, i.e. it may consist of only identity element.

Action on the displacement pattern $\boldsymbol{\delta}_j$ by those elements of the parent group $G_0$ that do not belong to the group $G_j$ generates the so-called "dynamical domains" of the considered dynamical regime [5-7] (this term is similar to that of the theory of structural phase transitions). All physical properties of such domains are identical.

As mentioned, the Rosenberg NNM represents a one-parametric dynamical regime. On the other hand, in nonlinear Hamiltonian system with a discrete symmetry group $G_0$, there can exist many exact multi-parametric regimes, determined by different symmetry groups $G_j$ ($G_j \subseteq G_0$). Namely these dynamical regimes represent bushes of NNMs.

---

[*] Here we do not discuss the more complex situation where the symmetry of the bush is lower than symmetry of all its modes [23, 24]. In this case, the group of the bush is the intersection of all the symmetry groups of individual NNMs entering the bush.

According to the above-said,

$$G_j \boldsymbol{\delta}_j = \boldsymbol{\delta}_j. \qquad (2)$$

This means that the displacement pattern $\boldsymbol{\delta}_j$ must be invariant under action of all elements of the given subgroup $G_j$ of the parent group $G_0$. The displacement pattern $\boldsymbol{\delta}_j$ can be written as a sum of contributions from different irreps of the group $G_0$:

$$\boldsymbol{\delta}_j = \sum_i \boldsymbol{\delta}(\Gamma_i) \qquad (3)$$

It was proved in the framework of the bush theory that one can obtain from (2) the following invariant relations for individual irreps $\Gamma_i$:

$$(\Gamma_i \downarrow G_j) c_i = c_i \qquad (4)$$

for all $\Gamma_i$ of the group $G_j$. Here $\Gamma_i \downarrow G_j$ is the *restriction* of the irrep $\Gamma_i$ of the group $G_0$ on its subgroup $G_j$[*].

Vector satisfying the relation (4) we call an *invariant* vector of the irrep $\Gamma_i$. In general, invariant vector of a given irrep depends on a number of arbitrary parameters that we denote by the letters *a, b, c, d*, etc. To simplify the form of our tables, we denote the above arbitrary parameters for different $\Gamma_i$ by the same letters, i.e., by default, we consider them to be different for different irreps.

We begin construction of the bushes of NNMs by looking at each irrep, considering it as a *root* representation, and find all its non-equivalent invariant vectors (see Appendix). In such a way, we can select all subgroups $G_j$ of the group $G_0$ associated with the set of invariant vectors of this irrep [28].

Then we fix a given $\Gamma_i$, one of its invariant vector $c_j$ and look again at all the irreps of the group $G_0$ for finding those to which the *secondary* modes of the bush can correspond (every bush is determined by the pair $\Gamma_i$ and $c_j$). For each restriction $\Gamma_i \downarrow G_j$, we find the invariant vector by solving the system of linear algebraic equations (4). As a result, we obtain the "complete condensate of the order parameters" [24, 29, 30], which defines the set of arbitrary coefficients entering the linear combination of basis vectors of the irrep $\Gamma_i$. In the case of vibrational bushes these basis vectors must be constructed in the space of all atomic displacements. This procedure allows us to determine the explicit form of the displacement pattern $\boldsymbol{\delta}_j$, which is invariant with respect to the selected group $G_j \subseteq G_0$.

---

[*] This restriction consists of all matrices of the irrep $\Gamma_i$ of the group $G_0$, which correspond to the symmetry elements of the group $G_j$ only ($G_j \subset G_0$).

Let us demonstrate the above-said using a concrete example. We consider a four-dimensional vibrational bush *B[Cmm2]* in graphene with parent symmetry group $G_0 = P6mm$.

One of the irreps of the group $G_0 = P6mm$ is three-dimensional representation $\Gamma_i = \Gamma_{12-2}$ (some comments on designations of the irreps of space groups are given below in Sec. 4). Considering this representation as a root irrep, we can find for its four non-equivalent invariant vectors: $(a,a,a), (a,a,0), (a,0,0), (a,b,c)$. The space groups selected by these vectors are *P6, Cmm2, Pba2* and *P112*, respectively.

Let us consider the invariant vector $(a, a, 0)$. It corresponds to the *root mode* of the form:

$$\boldsymbol{\delta}[\Gamma_{12-2}, (a,a,0)] = a\boldsymbol{\varphi}_1 + a\boldsymbol{\varphi}_2 + 0\boldsymbol{\varphi}_3$$

where $\varphi_i$ ($i = 1..3$) are basis vectors of the irreps $\Gamma_{12-2}$ which are constructed in the space of all atom displacements. These 16-dimensional vectors are presented in Table 1. For each atom we give its displacements along X and Y axis of hexagonal coordinate system. The numbering of the graphene atoms is given in Fig. 1.

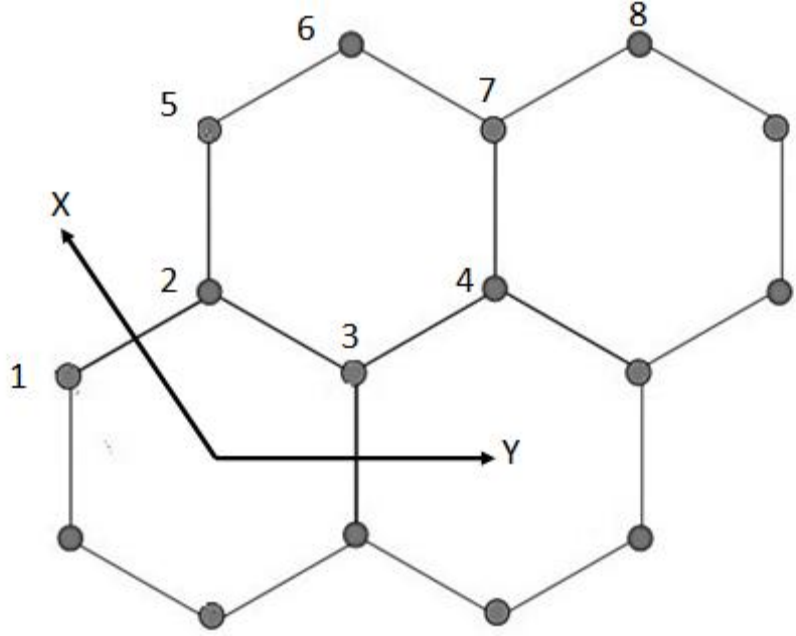

Fig. 1. Fragment of the graphene lattice with atomic numbering used in displacement patterns presented in Table 1 and Eq. (5).

Table 1. Displacement patterns corresponding to the basis vectors of irreducible representations of the graphene symmetry group $G_0 = P6mm$

| irrep | atom num. / basis vector | displacement pattern | | | | | | | | | | | | | | | |
|---|---|---|---|---|---|---|---|---|---|---|---|---|---|---|---|---|---|
| | | 1 | | 2 | | 3 | | 4 | | 5 | | 6 | | 7 | | 8 | |
| 12-2 | $\varphi_1$ | 1 | 0 | 1 | 0 | 1 | 0 | 1 | 0 | -1 | 0 | -1 | 0 | -1 | 0 | -1 | 0 |
| | $\varphi_2$ | -1 | 1 | 1 | -1 | 1 | -1 | -1 | 1 | 1 | -1 | -1 | 1 | -1 | 1 | 1 | -1 |
| | $\varphi_3$ | 0 | 1 | 0 | 1 | 0 | -1 | 0 | -1 | 0 | 1 | 0 | 1 | 0 | -1 | 0 | -1 |
| 12-1 | $\psi_1$ | 1 | -2 | 1 | -2 | 1 | -2 | 1 | -2 | -1 | 2 | -1 | 2 | -1 | 2 | -1 | 2 |
| | $\psi_2$ | 1 | 1 | -1 | -1 | -1 | -1 | 1 | 1 | -1 | -1 | 1 | 1 | 1 | 1 | -1 | -1 |
| | $\psi_3$ | 2 | -1 | 2 | -1 | -2 | 1 | -2 | 1 | 2 | -1 | 2 | -1 | -2 | 1 | -2 | 1 |
| 16-6 | $\chi_1$ | 0 | 1 | 0 | -1 | 0 | 1 | 0 | -1 | 0 | 1 | 0 | -1 | 0 | 1 | 0 | -1 |
| | $\chi_2$ | $\frac{\sqrt{3}}{2}$ | 0 | $-\frac{\sqrt{3}}{2}$ | 0 | $\frac{\sqrt{3}}{2}$ | 0 | $-\frac{\sqrt{3}}{2}$ | 0 | $\frac{\sqrt{3}}{2}$ | 0 | $-\frac{\sqrt{3}}{2}$ | 0 | $\frac{\sqrt{3}}{2}$ | 0 | $-\frac{\sqrt{3}}{2}$ | 0 |

Now let us consider all *secondary* modes corresponding to the above root mode. With the aid of the group-theoretical methods, we can find that only two secondary vibrational modes correspond to this root mode. They belong to the two-dimensional irrep $\Gamma_{16-6}$ and to three-dimensional irrep $\Gamma_{12-1}$. The displacement patterns of these modes can be written in the form:

$$\delta[\Gamma_{12-1}, (a, -a, b)] = a\psi_1 - a\psi_2 + b\psi_3,$$

$$\delta\left[\Gamma_{16-6}, \left(a, -\frac{\sqrt{3}}{2}a\right)\right] = a\chi_1 - \frac{\sqrt{3}}{2}a\chi_2.$$

Note, that the term "mode" we use for any linear combination of the basic vectors of an irreducible representation which may depends on a number of arbitrary parameters (each mode corresponds to a certain symmetry group of the crystal vibrational state). In the above example, the mode corresponding to the irrep $\Gamma_{16-6}$ depends on only one parameter ($a$), while the mode corresponding to the irrep $\Gamma_{12-1}$ depends on two parameters ($a$, $b$). Let us remind that letters "$a$" entering the displacement patterns of different irreps possess *different values*!

Basis vectors of the above irreps $\Gamma_{16-6}$ and $\Gamma_{12-1}$, generating the secondary modes, are given in Table 1. Thus, we obtain the final form of the considered vibrational bush *B[Cmm2]*:

$$\delta\{B[Cmm2]\} = (a\varphi_1 + a\varphi_2) + (a\psi_1 - a\psi_2 + b\psi_3) + \left(a\chi_1 - \frac{\sqrt{3}}{2}a\chi_2\right).$$

Here, the contributions from individual irreps are enclosed in brackets with *different* values of parameter $a$ for different representations.

The following displacement pattern corresponds to the above bush:

$$\delta\{B[Cmm2]\} = (A, B|2C, -C|2D, -D|-A, A+B|A, -A-B|-2D, D|-2C, C|-A, -B). \quad (5)$$

Here, letters *A, B, C, D* denote four arbitrary parameters which determine the instant $x_k$ and $y_k$ displacements of all 8 graphene atoms in 2x2 graphene primitive cell of the *vibrational* state.

### 4. Group-theoretical results on nonlinear vibrations of the graphene lattice

As was already discussed, we consider in-plane nonlinear vibrations in a single-layer graphene possessing the space symmetry group $G_0 = P6mm$ (the graphene may be subjected to the uniform stretching because it conserves the symmetry $P6mm$).

In some sense, the bush theory is a generalization of the theory of the complete condensate of order parameters which was developed earlier in the framework of studying structural phase transitions [24, 29, 30] (in turn, it is a generalization of the Landau theory of the second order phase transitions [31]).

Particularly important role in the Landau theory play the so-called *commensurate transitions* [31]. For such transitions the primitive cell volume in the low-symmetry phase is integer times larger than that of the high-symmetry phase. It was proved in the framework of this theory that for stability of the low-symmetry phase against long-wave fluctuations, the so-called soft Lifshitz condition is necessary [31]. According to this condition, the structure of the low-symmetry phase must correspond to an irreducible representation of the group $G_0$, which is associated with one of the high-symmetry points in the Brillouin zone[*].

Considering nonlinear oscillations of the crystal lattice, we also deal with lowering of the symmetry from the group $G_0$ of the equilibrium state to the group $G_j$ of vibrational state ($G_0 \to G_j$). In this connection, analyzing nonlinear oscillations in graphene, we also consider only those vibrational states, which are described by irreps corresponding to the points of high symmetry in the Brillouin zone. As a consequence, the primitive cell of such vibrational state (we call it "supercell") is the integer times larger than the primitive cell of the graphene in equilibrium.

It is well known, that irreducible representations of space groups are determined by two indices [32]. They are wave vector in Brillouin zone and the number of irrep of the symmetry group of this vector. There are three high-symmetry points in Brillouin zone of the space group P6mm [32]. They determined by the wave vectors $k_{16} = (0,0), k_{12} = \left(\frac{1}{2}, 0\right)$ and $k_{13} = \left(\frac{1}{3}, \frac{1}{3}\right)$. Four one-dimensional and two two-dimensional irreps correspond to the vector $k_{16}$, four three-dimensional irreps correspond to the vector $k_{12}$, two two-dimensional and one four-dimensional irreps are associated with the vector $k_{13}$.

Let us present some of our group-theoretical results on the bushes of NNMs in graphene (computational details will be published elsewhere).

---

[*] Irreps, associated with lines and planes of symmetry in the Brillouin zone, generate incommensurate structures.

In graphene structure, there exist 37 different vibrational bushes corresponding to high-symmetry points in the Brillouin zone. There are 4 one-dimensional, 14 two-dimensional, 1 three-dimensional and 6 four-dimensional among them. The atomic displacement patterns for one-dimensional bushes (they represent symmetry determined Rosenberg nonlinear normal modes) are shown in Fig. 2-5 (supercells of the vibrational states are depicted by dotted lines). Let us note that two *different* one-dimensional bushes possess the same group *P6mm*. For two-dimensional bushes we restrict ourselves only by presenting information in Table 2. In this table, the complete condensates for these bushes are given. Note, that irreps 16-1, 16-2, 16-3 and 16-4 do not enter into the mechanical representation of the group $G_0 = P6mm$ and, therefore, they cannot contribute to the bush displacement patterns. Only 3 two-dimensional bushes are associated with individual irreps (entries 3, 11, 14 in Table 2 correspond to these bushes), while all other bushes in these table are associated with two different irreps.

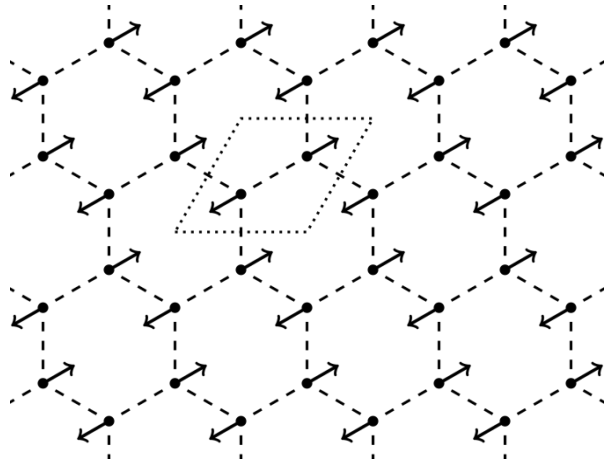

Fig. 2. One-dimensional bush with space group Cmm2.

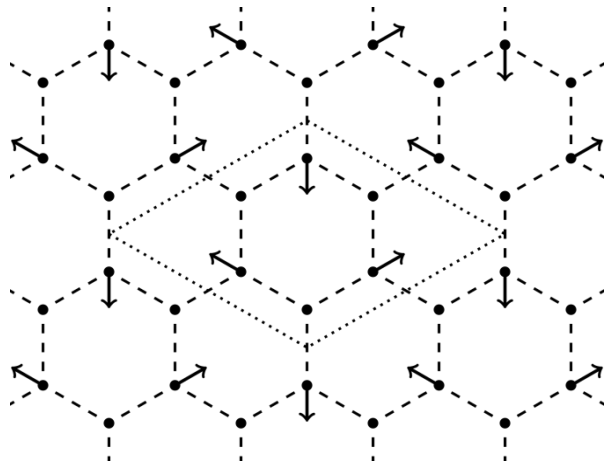

Fig. 3. One-dimensional bush with space group P31m.

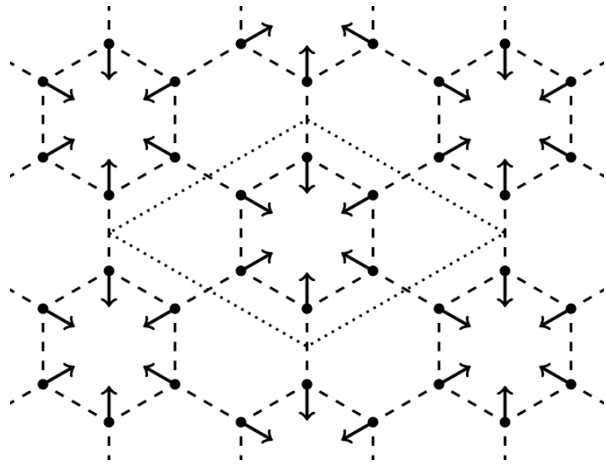

Fig. 4. One-dimensional bush with space group P6mm.

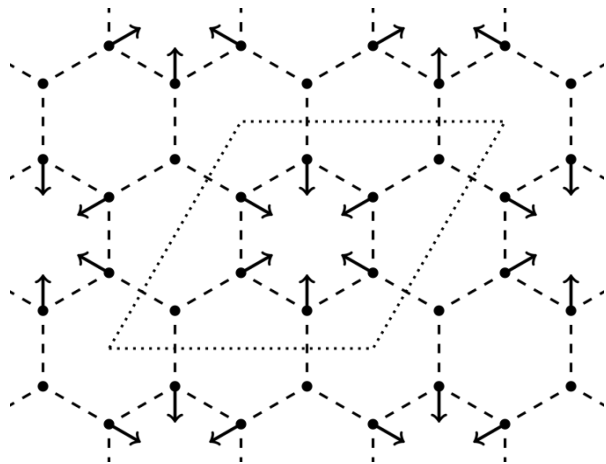

Fig. 5. One-dimensional bush with space group P6mm.

Table 2. Complete condensates of order parameters corresponding to two-dimensional vibrational bushes. Root modes are highlighted by gray color

| # | Space group | Irreducible representation | | | | | | | | | | | |
|---|---|---|---|---|---|---|---|---|---|---|---|---|---|
| | | 16-1 | 16-2 | 16-3 | 16-4 | 16-5 | 16-6 | 12-1 | 12-2 | 12-3 | 12-4 | 13-1 | 13-2 | 13-3 |
| 1 | c1m1 | *a* | | | *a* | 0,*a* | *a*,0 | | | | | | | |
| 2 | c1m1 | *a* | | *a* | | *a*,0 | *a*,0 | | | | | | | |
| 3 | p112 | *a* | *a* | | | | *a,b* | | | | | | | |
| 4 | pmm2 | *a* | | | | | *a*,0 | *a*,0,0 | | | | | | |
| 5 | p6 | *a* | *a* | | | | | *a,a,a* | *a,a,a* | | | | | |
| 6 | pba2 | *a* | | | | | *a*,0 | | *a*,0,0 | | | | | |
| 7 | p3m1 | *a* | | *a* | | | | *a,a,a* | | *a,a,a* | | | | |
| 8 | pbm2 | *a* | | | | | *a*,0 | | | *a*,0,0 | | | | |
| 9 | p31m | *a* | | | *a* | | | *a,a,a* | | | *a,a,a* | | | |
| 10 | pbm2 | *a* | | | | | *a*,0 | | | | *a*,0,0 | | | |
| 11 | p3m1 | *a* | | | *a* | | | | | | | *a,b* | | |
| 12 | p6 | *a* | *a* | | | | | | | | | *a*,0 | *a*,0 | |
| 13 | p31m | *a* | | *a* | | | | | | | | *a*,0 | 0,*a* | |
| 14 | p3 | *a* | *a* | *a* | *a* | | | | | | | | | *a,b,-b,a* |

In the second column of Table 2, we give *space groups* of the bushes. As was already discussed, they represent the symmetry groups of the atomic displacement patterns corresponding to the given bushes. As was discussed in Sec. 3, the explicit form of these displacement patterns can be obtained by combining the results from Table 2 with basis vectors of the irreps of the group $G_0 = P6mm$ constructed in the space of all atomic displacements.

Note that, as well as in the case of one-dimensional bushes, some identical symbols of space groups correspond to different bushes in Table 2. One of the causes of this phenomenon is that these groups, being subgroups of $G_0 = P6mm$, possess different embeddings in this parent symmetry group. For example, the space group *C1m1* for the bush #1 and the same group for the bush #2 correspond to *different* orientation of the mirror planes orthogonal to the graphene sheet (in the atomic chains parallel to these planes the nearest carbon atoms are spaced from each other at different distances).

## 4. Conclusion

In this paper, we present only some results of the group-theoretical analysis of in-plane vibrational states in monolayer graphene with the symmetry group *P6mm* (more detailed information will be published elsewhere). These results were obtained with the aid of the general theory of bushes of nonlinear normal modes, which was developed in [5-7]. We have found that only four one-dimensional bushes can exist in graphene. They represent symmetry determined Rosenberg nonlinear normal modes and describe periodic vibrations of the graphene lattices. We have also found that 14 two-dimensional, 1 three-dimensional and 6 four-dimensional vibrational bushes can exist in the considered system (they represent *quasi-periodic* dynamical regimes).

Let us outline several directions of further research of nonlinear oscillations in graphene.

1. The adequacy of the group-theoretical analysis of the bushes of NNMs can be verified with the aid of ab initio calculations based on the density functional theory (the similar study for the molecule $SF_6$ was published in [14]). Such study allows also to find Fourier spectra of the corresponding vibrational regimes, to analyze dynamics of the system dipole moment, etc.

2. It is necessary to investigate the influence of temperature on the dynamics of the vibrational bushes in graphene.

3. Possible ways of excitation of bushes of nonlinear normal modes should be investigated. It seems that, in principle, this can be done by using radiation of two lasers with appropriate polarization at close frequencies with the aim at coincidence of their beat frequency with the frequency of the nonlinear oscillations in graphene (a similar study one can see in [33]).

**Acknowledgments**

*Authors are grateful for financial support: G.M. Chechin and D.S. Ryabov to the Russian Science Foundation (Grant No. 14‑13‑00982), S.A. Shcherbinin to the Southern Federal University (Internal grant No. 213.01-07.2014/11ПЧВГ).*

**Appendix**

As was already mentioned, the theory of bushes of NNMs was developed as a generalization of the theory of complete condensates of order parameters for studying structural phase transitions [29, 30]. In the framework of this theory, the algorithm for finding all invariant vectors of irreducible representations of the group $G_0$ was developed in [28]. Each invariant vector $c_j$ selects the set of matrices of the given irrep $\Gamma_i$ which conserve it. Elements of the group $G_0$, corresponding to the above matrices, form a certain subgroup $G_j \supseteq G_0$. Let us note that to each element of the group $G_0$ corresponds only one matrix of $\Gamma_i$, while the same matrix can associate with different symmetry elements of the group $G_0$ because of homomorphism between symmetry group and its representation. The full set of all different matrices of the given irrep forms the *image* of this representation. Obviously, we can use only the image of $\Gamma_i$ when we search the set of its invariant vectors. Usually, among these invariant vectors there are many *equivalent*. Such vectors correspond to the *conjugate* subgroups in the group $G_0$ (invariant vectors are called equivalent, if they can be transform into each other at least by one of the matrices of the considered representation $\Gamma_i$). Equivalent invariant vectors correspond to the symmetry groups identical in the crystallographic sense (in turn, they correspond to the different domains of the same phase), while nonequivalent invariant vectors correspond to different phases.